\documentclass[10pt,twocolumn]{article}
\usepackage{amsmath}
\usepackage{amsfonts}
\usepackage{amssymb}
\usepackage{graphicx}
\usepackage{bm}
\title{\textbf{Absolute cross-sections of fragment negative ions in electron collisions with difluoromethane}}
\author{\textbf{Dipayan Chakraborty$^1$} and \textbf{Dhananjay Nandi$^2$}\\ Indian Institute of Science Education and Research Kolkata, Mohanpur 741246, India\\ \small{email:$^1$dc14rs010@iiserkol.ac.in, $^2$dhananjay@iiserkol.ac.in}}

\date{}
\begin{document}
\twocolumn[
  \begin{@twocolumnfalse}
    \maketitle
    \begin{abstract}
      Dissociative electron attachment (DEA) and ion-pair dissociation (IPD) processes of Difluoromethane (CH$_2$F$_2$) have been studied in the incident electron energy range 0 to 45 eV. Three different fragment anions (F$^-$, CHF$^-$ and F$_2^-$) are detected in the DEA range and two anions (F$^-$ and CHF$^-$) are detected in IPD range. Absolute cross-section of the F$^-$ fragment ion is measured for the first time. Three different resonances for both F$^-$ and CHF$^-$ ions and one single resonance peak for the F$_2^-$ ions are observed. Constant increase in ion counts above 8 eV incident electron energy indicates the involvement of IPD process. From the experimental observation, it is speculated that near 11 eV incident electron energy both DEA and IPD processes occur simultaneously. 
    \end{abstract}
  \end{@twocolumnfalse}
]
\section{Introduction}

Total elastic and inelastic electron scattering cross-section studies of fluoromethanes is a topic of interest these days \cite{christophorou}. Cross-section values of these molecules have a demand because of its application to plasma processing in the semiconductor industry. Besides this, Chlorofluorocarbons (CFC) and Hydrofluorocarbons (HFC) are the main reason behind the Ozone layer depletion due to their photolytic decomposition. However decomposition can also occur due to the low energy electron attachment process \cite{dea:illenberger}. So, it is absolutely necessary to have accurate electron attachment cross-section values of these molecules. Several studies are performed to address this problem \cite{stano,rawat,allan} but for Methylene fluoride (CH$_2$F$_2$) it is very rare.\\
Electronic structures and energies of the Fluoromethanes have been studied long ago by Brundle \emph{et al.} \cite{brundle}. In 1997, Tanaka \emph{et al.} \cite{tanaka} obtain their elastic differential cross sections below 100 eV incident electron energy, later the integral cross-section is also calculated \cite{tanaka1}. Later Tamio Nishimura \cite{nishimura} theoretically calculate the vibrationally elastic scattering cross section of CH$_2$F$_2$ with electron collision below 30 eV. All these studies are limited to only elastic electron scattering cross section of CH$_2$F$_2$ but the inelastic scattering processes like dissociative ionization (DI), DEA and IPD process of CH$_2$F$_2$ with electron collisions are not studied in detail. In 1998, Motlagh and Moore \cite{motlagh} studied the electron impact DI process of CH$_2$F$_2$ molecule up to 500 eV incident electron energy range. Later Torres \emph{et. al} \cite{torres} have studied the same upto 100 eV energy range by using time-of-flight mass spectrometry method. The authors have discussed the appearance energy, absolute total and dissociative ionization cross sections and the corresponding kinetic energy of the fragment ions in details. The other inelastic scattering processes like DEA and IPD are still ignored for CH$_2$F$_2$ though it is equally important like other molecules in fluoromethane group. As per authors concern the only reported studies of  DEA and electron impact IPD process of CH$_2$F$_2$ is done by Scheuremann \emph{et al.} \cite{Illenberger}. In this article, the authors have observed the formation of F$^-$ and CHF$^-$ fragment ions up to 20 eV incident electron energy range. Excitation function of the fragment anions beyond 6 eV incident electron energy is reported, although the corresponding absolute cross-section values are not measured. In the present article, DEA and IPD processes of CH$_2$F$_2$ are studied from 0 to 45 eV incident electron energy, using an advance time of flight mass spectrometer (TOFMS) developed in our group \cite{rsi:dipayan}. Three different fragment anions are observed. Dissociation pathways and corresponding appearance energies of the fragment ions are discussed based on the experimental observations. One low energy temporary negative ion (TNI) state around 2 eV incident electron energy is observed for the first time followed by three higher energy TNI states, in agreement with the previous report \cite{Illenberger}. Absolute DEA and IPD cross-sections of the negative ions are measured in the above mentioned energy range.

\section{Instrumentation}

Details of the experimental setup and the measurement procedure is described elsewhere \cite{rsi:dipayan}. Here the measurement technique is discussed briefly. Basic theme of the experiment is magnetically collimated pulsed electron beam with 200 ns pulse width and 10 kHz repetition rate is interacted perpendicularly with an effusive molecular beam produced through a needle of diameter 1 mm. Tip of the needle was kept 4 mm away from the interaction region. Negative ions formed in the interaction region are guided through a spectrometer and collected by the detector. The energy of the emitted electrons is controlled by an external power supply which is connected with the electron gun filament. Filament current is provided by a constant current supply and the electrons are emitted via thermionic emission process. The electron beam current has been measured by using a Faraday cup, placed opposite to the electron gun in the interaction region. Time-averaged electron beam current during the measurement was around 3 nA. Two magnetic coils in Helmholtz configuration are used to collimate the electron beam. Typical strength of the magnetic field is 30 Gauss. Axis of the spectrometer is situated perpendicular to both electron beam and molecular beam. Spectrometer contains a pusher plate, a puller plate, three lens electrodes, a drift tube and a mesh grid. The electron-molecule interaction occurs in between pusher and puller plates. These pusher and puller plates consist of wire mesh of 90\% transmission efficiency to avoid field penetration into the interaction region. After the puller plate, three lens electrodes are placed in Einzel lens configuration to focus the negative ions. Applied voltage to the three electrodes are 90, 1030 and 90 volts respectively. In order to increase the mass resolution of the spectrometer, one field free drift tube is placed after the lens electrodes. At the end of the drift tube, one cap electrode with wire mesh is placed to avoid field penetration from the detector. Both the drift tube and the cap electrode is biased to 1590 V. After the drift tube, MCP based detector is placed to collect the negative ions. The detector consists of two micro channel plates (MCP) in Chevron configuration, along with one collector plate. The TOF of the detected ions are determined from the back MCP signal. Experiments were performed under ultra high vacuum conditions with base pressure $10^{-9}$mbar and with 99.9\% pure commercially available CH$_2$F$_2$ gas.\\
The absolute cross-section of the F$^-$ fragment anion has been measured by using relative flow technique (RFT) \cite{EK:RFT,orient:RFT,srivastava}. RFT is basically a calibration procedure where one just needs to compare the relative intensities of the species of interest with a standard species of known cross section, by keeping the other experimental conditions unchanged. For example, in the present case, the absolute cross section for the formation of F$^-$ fragment ions from CH$_2$F$_2$ can be determined by using the dissociative electron attachment (DEA) cross section \cite{ref:rapp} of O$^-$ ion from O$_2$ by using the equation as
\begin{equation}
\begin{split}
\sigma(\text{F}^-/\text{CH}_2\text{F}_2)=\sigma(\text{O}^-/\text{O}_2) \frac{N(\text{F}^-)}{N(\text{O}^-)}\frac{I_e(\text{O}_2)}{I_e(\text{CH}_2\text{F}_2)} \times \\
\left(\frac{M_{\text{O}_2}}{M_{\text{CH}_2\text{F}_2}}\right)^{1/2}\frac{F_{\text{O}_2}}{F_{\text{CH}_2\text{F}_2}}\frac{K(\text{O}^-)}{K(\text{F}^-)}
\end{split}.
\end{equation}
Here $N$ is the number of fragment ions collected for a fixed time, $F$ is the flow rate of the corresponding gases, $I_e$ is the time average electron beam current, $M$ is the molecular weight of the parent molecules, $K$ is the detection efficiency and $\sigma$ is the absolute cross section. All these factors and their contributions in the overall measurements are discussed in details in the previous study \cite{rsi:dipayan}.\\
\begin{figure}
\centering
\includegraphics[scale=.44]{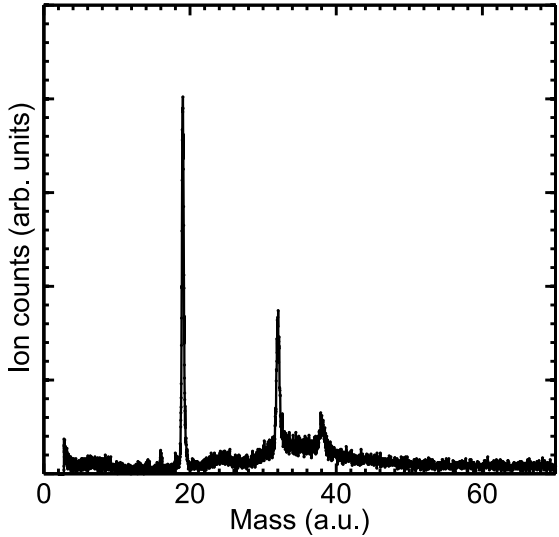}
\caption{Mass spectra of the CH$_2$F$_2$ at 11 eV incident electron energy. Three different masses F$^-$, CHF$^-$ and F$_2^-$ are shown respectively.} \label{mass_spec}
\end{figure}

\section{Results and Discussion}
\begin{figure}
\centering
\includegraphics[scale=.43]{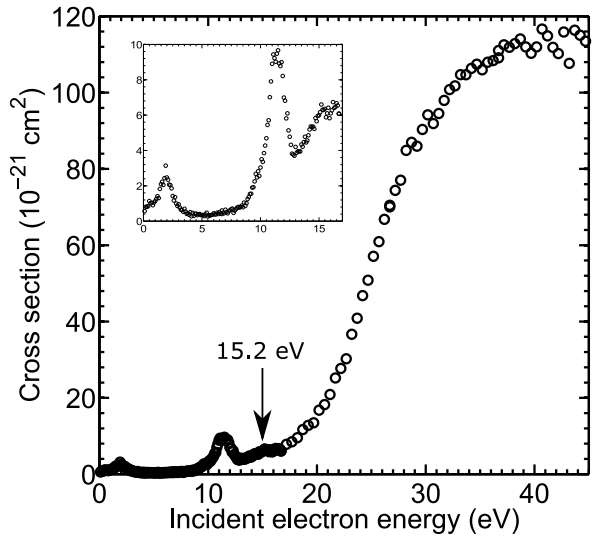}
\caption{Ion yield curve for F$^-$ ion. Two lower energy peaks at 1.9 eV and 11.4 eV are observed. Two small humps at 10.1 eV and 15.2 eV are shown by the arrow.} \label{F_ionyield}
\end{figure}

When low energy electrons are collided with the neutral molecules, temporary negative ion (TNI) states (CH$_2$F$_2$)$^{*-}$ are formed via dissociative electron attachment (DEA) process. This TNI further decays via three possible dissociation channels
\begin{equation}
\text{CH}_2\text{F}_2 +\text{e}^-\rightarrow(\text{CH}_2\text{F}_2)^{-*}\rightarrow \left\{
\begin{array}{c}
\text{F}^- + \text{CH}_2\text{F}\\
\text{CHF}^- + \text{HF}\\
\text{F}_2^- + \text{CH}_2
\end{array}
\right.
\end{equation}
Here F$^-$ channel is a simple bond cleavage whereas, CHF$^-$ and F$_2^-$ channels are associated with the rearrangements in the TNI. Fig. \ref{mass_spec} represents the mass spectra of CH$_2$F$_2$ molecule obtained at 11 eV incident electron energy. Here X-axis is calibrated in the mass unit which reveals that, in this energy region three different fragment ions F$^-$, CHF$^-$ and F$_2^-$ are formed. In the previous experimental study, first two channels (F$^-$ and CHF$^-$) were observed \cite{Illenberger} but the third channel(F$_2^-)$ is observed for the first time. Fig. \ref{F_ionyield}, \ref{CHF_ionyield} and \ref{F2_ionyield} are the corresponding excitation function of the three fragment ions. From those excitation functions one can observe that for F$^-$ ions, one low energy peak around 2 eV followed by two higher energy peaks around 11 eV and 15.2 eV are present. Whereas for CHF$^-$ ions, along with 2 eV peak, only one broad peak around 11 eV is present. With close inspection one small hump near 10 eV incident electron energy for F$^-$ ion can also be observed. But for F$_2^-$ ions only one broad peak near 2 eV incident electron energy is observed. From these observations one can conclude that in the Franck-Condon (FC) transition region of neutral CH$_2$F$_2$ molecule, temporary negative ion (TNI) states are present around those energies. 
\subsection{Resonance peak around 2 eV}
Presence of lower energy peak due to DEA of chlorofluorocarbon is quite natural and has been observed in several studies \cite{langer}. In 1992 Modelli \emph{et. al.} \cite{modelli} studied the electron attachment to the halomethanes via electron transmission spectroscopy and observed low energy electrons are attached with the halomethanes though for fluoromethanes they were unable to measure any low energy ($< 6$ eV) resonances. In 2000 Langer \emph{et. al.} measured the negative ion formation due to low energy electron collision to CF$_2$Cl$_2$. In this study, the authors observed low energy NIR state which dissociates via different fragment negative ions. They conclude that the NIR state, which dissociates via Cl$^-$ negative ion formation, is formed due to Shape Resonance, where the incoming electron is occupying the molecular orbital with $\sigma^*$(C-Cl) character. However they are unable to comment on the nature of the NIR state which dissociates via F$^-$ dissociation channel.
According to author's concern for CH$_2$F$_2$ molecule, below 6 eV incident electron energy region no experimental study has been done till date.\\ 
Within the FC transition window the low energy electron attachment to the ground state CH$_2$F$_2$ creates the TNI states which further dissociates to the negative fragment ions via corresponding dissociation channels. After formation of the TNI, there will be two competing channels. One is the dissociation of the TNI via one negative and other neutral fragment, other is the auto detachment (AD) where the TNI ejects the electron and back to the parent neutral molecule (may be with  vibrationally excited states). The life time of the TNI actually determines the cross-section or the probability of negative ion formation. If the life time of the TNI is subsequently high so that it can cross the critical distance R$_{\text{c}}$, then the DEA cross section will be high otherwise AD of the TNI occurs. In the present context, measured DEA cross section is around $10^{-21}$cm$^2$ for the F$^-$ channel. For CHF$^-$ and F$_2^-$ channels the cross-section is too low to calculate any reliable absolute value, thus only differential values are reported.\\

\subsection{Resonances above 6 eV}
\begin{figure}
\centering
\includegraphics[scale=.435]{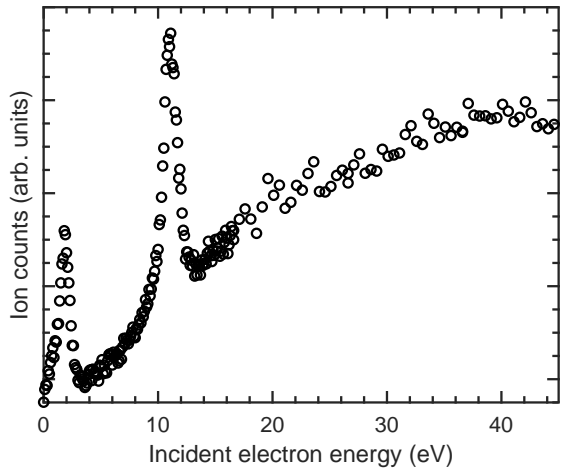}
\caption{Ion yield curve for CHF$^-$ ion. Two lower energy peaks at 1.9 eV and 11.4 eV are observed.} \label{CHF_ionyield}
\end{figure}
\begin{figure}
\centering
\includegraphics[scale=.44]{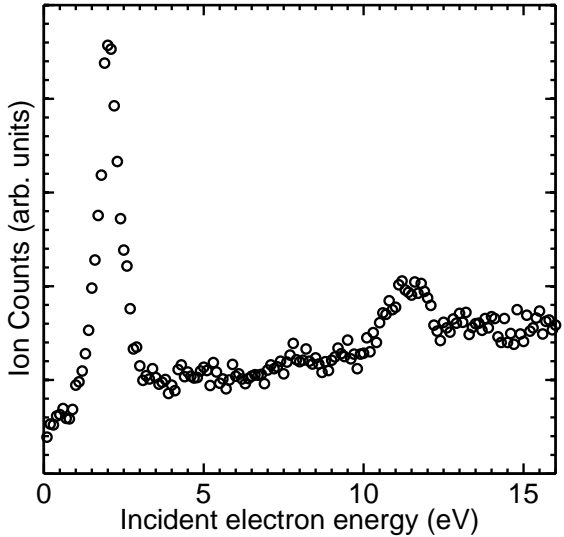}
\caption{Ion yield curve for F$_2^-$ ion. One lower energy peak around 1.9 eV electron energy is prominent and one small hump around 11.4 eV is shown.} \label{F2_ionyield}
\end{figure}

In the measured excitation function of the three channels (Fig. \ref{F_ionyield}, \ref{CHF_ionyield} and \ref{F2_ionyield}) one resonance peak near 11 eV is observed. Two small humps near 10 eV and 15.2 eV incident electron energy for F$^-$ ions are also observed. This higher energy resonance peaks can be described from previous experimental as well as theoretical studies. Using electron transmission spectroscopy (ETS) and multiple scattering X$\alpha$ (MS-X$\alpha$) bound state calculations, Modellii \emph{et al.} \cite{modelli} studied the electron attachment to halomethanes. In this study, the experimentally obtained electron attachment energies and theoretically calculated values match nicely. Using MS-X$\alpha$ values the authors predict that in CH$_2$F$_2$ molecule, two broad $\sigma^*$ resonances with symmetries b$_2$ and a$_1$ with comparable intensities around 10 eV are responsible for the electron attachment cross-section. In the present study, clear signature of the TNI states around those above-mentioned energy ranges are observed. Two peaks as mentioned in the theoretical study is not possible to observe separately due to poor electron gun resolution. Besides these two DEA resonances, one peak near 15.2 eV of the F$^-$ ion yield curve has been observed, but due to unavailability of any theoretical studies we are unable to make any comment on that. However, the presence of 15.2 eV peak in the excitation function of F$^-$ ion has been observed previously by Scheuremann \emph{et al.} \cite{Illenberger}.\\
\subsection{Ion-pair dissociation}
With close inspection of the excitation function, one can observe the cross-sections of F$^-$, CHF$^-$ ions gradually increasing beyond 8 eV incident electron energy. This clearly indicates that the ion-pair dissociation (IPD) process starts around this energy. Same behaviour is observed in the previous measurement \cite{Illenberger}. Unlike DEA process, in IPD process the electrons are not captured by the molecule. Here the incident electron transfers some energy to the molecule and excites it to the ion-pair state that eventually dissociates into an anion and a cation. This ion-pair dissociation is possible as long as the energy of the incident electron is equal to or more than the excitation energy required because the extra energy is always carried out by the outgoing electron \cite{dipayan:co}. The IPD process in the present experiment can be expressed as
\begin{equation}
\text{CH}_2\text{F}_2 +\text{e}^- \rightarrow \left\{
\begin{array}{c}
\text{F}^- + \text{CH}_2\text{F}^+ + \text{e}^-\\
\text{CHF}^- + \text{HF}^+ + \text{e}^- 
\end{array}
\right.
\end{equation}
Due to low cross-section, we are unable to comment anything about the F$_2^-$ channel. For F$^-$ channel the cross section is quite large in the IPD range compared to DEA range whereas, for CHF$^-$ channel the cross-sections in the two regions are comparable. By looking at the experimental results and the present understanding it is speculated that in the FC transition region the TNI state lies above the ion-pair state. As a result, around 11 eV incident electron energy both the DEA and IPD processes occurred simultaneously.\\
Obtained absolute cross section value of F$^-$ ion is listed in Table \ref{table1}. Extreme care has been taken to confirm that these peaks are not coming from impurities. The mass resolution of the spectrometer is high enough that one can separate the mass difference of 1 amu within this range with kinetic energy of fragments up to 5 eV. As mentioned earlier the experiments have been performed with 99.9 \% pure CH$_2$F$_2$ gas and the chamber is kept in ultra high vacuum ($10^{-9}$ mbar) for more than one week before the experiment. So we confirmed the error due to the impurity is negligible here. As the cross-section is very low, there is a probability that the peaks in the cross-section curve are present due to the collision with secondary electrons \cite{chantry}. This can be identified by non linear pressure dependence of the peak intensities. To avoid this secondary effects we have used very low background pressure ($10^{-7}$mbar). Thus we confirm the peaks appearing in both the cross-section curves are completely due to the dissociation of CH$_2$F$_2$ with electron collision. It is to be mentioned here that the overall uncertainty in our measurement is within 15\% \cite{rsi:dipayan}.
\begin{center}
\begin{table*}
\caption{Absolute cross section for the formation of F$^-$ and CHF$^-$ ions due to electron collisions with ground state CH$_2$F$_2$ molecule. The peak positions are in units of eV and the cross sections ($\sigma$) are in units of $10^{-20}$ cm$^2$.}
\begin{tabular}{c c c } \\ \hline
Ion & Peak position (eV) & Peak cross section ( $\times 10^{-21}$ cm$^2$) \\ 
 & & RFT \\ \hline
F$^-$/CH$_2$F$_2$ & 1.9 & 3.14 \\
 & 11.4 & 9.75 \\
  & 15.2 & 6.64 \\ \hline
  \end{tabular} \label{table1}
 \end{table*}  
 \end{center}

\section{Conclusion}
The absolute cross-section of F$^-$ ions due to electron collision with CH$_2$F$_2$ has been measured for the first time from 0 to 45 eV electron energy range by using relative flow technique (RFT). One low energy peak near 2 eV followed by higher energy peaks near 10, 11 and 15.2 eV has been observed. These peaks are the signature of dissociative electron attachment process. The ground state symmetry of CH$_2$F$_2$ is C$_{2v}$ \cite{nishimura}. After colliding with lower energetic electrons ($< 15$ eV), it forms temporary negative ion (TNI) state (CH$_2$F$_2$)$^{-*}$ and dissociates via three different dissociation channels producing F$^-$, CHF$^-$ and F$_2^-$ fragment anions. Depending upon the energy and possible symmetries of the remaining neutral fragments, threshold of different dissociation channels may vary. More than one different symmetries of (CH$_2$F$_2$)$^{*-}$ may involve in this dissociation process. In the present measurements, we are unable to comment on the possible symmetry of the TNI states. It was theoretically predicted that two negative ion resonant states of symmetries b$_2$ and a$_1$ are present around 10 eV incident electron energy. In the present measurements the presence of TNI states around 10 eV electron energy range is experimentally verified. Though due to poor electron energy resolution, we are unable to separate them out. The constant increase in the cross-section curve indicates that beyond 8 eV incident electron energy IPD process starts. More than one different symmetries of ion-pair states may involve in the process. From the experimental observation it is speculated that near 11 eV electron energy, both DEA and DD processes occur simultaneously.

\section{Acknowledgements}
D. N. gratefully acknowledges the partial financial support from ``Indian National Science Academy (INSA)" under project No. ``SP/YSP/80/2013/734" and from ``Science and Engineering Research Board (SERB)'' under the project No. ``EMR/2014/000457''. DC is thankful to IISER Kolkata for providing research fellowship.

\bibliographystyle{h-physrev}
\bibliography{ch2f2bib.bib}

\end{document}